\documentclass[prd,aps,floats,twocolumn]{revtex4}
\usepackage{latexsym}
\usepackage{amssymb}
\usepackage{amsmath}

\usepackage{epsfig}
\newcommand{\etal}{{\it et al.}}

\newcommand{\TD}{\ensuremath{D}}

\newcommand{\grad}{\ensuremath{\vec{\nabla}}}

\newcommand{\invq}{\ensuremath{{({q}^{-1})}}}
\newcommand{\rhob}{\ensuremath{\bar{\rho}}}
\newcommand{\E}{\ensuremath{{E}}}
\newcommand{\Ll}{\ensuremath{{\ell}}} 
\newcommand{\Pb}{\ensuremath{\bar{P}}}

\begin{document}

\preprint{\
\begin{tabular}{rr}
&
\end{tabular}
}
\title{Eddington-Born-Infeld gravity and the large scale structure of the Universe}
\author{M.~Ba\~nados$^{1}$,
P.G.~Ferreira$^{2}$ and C.~Skordis$^{3}$}
%
\affiliation{$^{1}$Departamento de F\'{\i}sica,
 P. Universidad Cat\'{o}lica de Chile, Casilla 306, Santiago 22,Chile.\\
$^2$Astrophysics, University of Oxford, Denys Wilkinson Building, Keble Road, Oxford OX1 3RH, UK\\
$^3$Perimeter Institute, Waterloo, Ontario N2L 2Y5, Canada.\\\
}

\begin{abstract}
It has been argued that a Universe governed by Eddington-Born-Infeld gravity can be compatible with current cosmological constraints.
The extra fields introduced in this theory can behave both as dark matter and dark energy, unifying the dark sector
in one coherent framework. We show the various roles the extra fields can play in the expansion of the Universe and
study the evolution of linear perturbations in the various regimes. We find that, as a unified theory of the dark sector, Eddington-Born-Infeld gravity
will lead to excessive fluctuations in the Cosmic Microwave Background on large scales. In the presence of a cosmological constant, however,
the extra fields can behave as a form of non-particulate dark matter and can lead to a cosmology which is entirely compatible with current
observations of large scale structure. We discuss the interpretation of this form of dark matter and how it can differ from standard, particulate dark matter.

\end{abstract}

\date{\today}
\pacs{PACS Numbers : }
\maketitle

\section{Introduction}
There is compelling evidence that baryonic matter in the presence of Einstein gravity does not suffice to describe
the Universe we live in. The most natural and popular suggestion is that we are surrounded by a sea of massive,
non-relativistic particles. Dubbed Cold Dark Matter (CDM), it can account for the dynamics of galaxies and clusters and
the large scale structure of the Universe. Furthermore it can arise in a plethora of extensions to the standard model
of particle physics and would currently exist as a thermal relic from hot era at early times \cite{peacock}.

Alternatives to the CDM scenario have been proposed. At the more extreme level, it has been suggested that Einstein
Gravity is modified, either through higher order corrections to the Einstein-Hilbert action, or through the addition of
new gravitational degrees of freedom that affect the relationship between the geometric and physical nature of the
space-time metric \cite{fofr,milgrom,teves,aether}. The less radical proposals typically involve replacing the Cold Dark Matter by some non-particulate
degree of freedom such as a scalar field or a fluid which has an effectively pressureless equation of state. Within this
class of models, there have been attempts at resolving {\it both} the dark matter problem and the dark energy problem. A
notable example is that of Chaplygin gas \cite{Kamenshchik-Moschella-Pasquier,Bertolami}.

A proposal has been put forward in \cite{banados1} from a different approach.  A theory for degenerate metrics is lacking and it was suggested that the solution would be the introduction of additional dynamical
degrees of freedom for the space-time connection \cite{banados1,banados3}. A candidate action, the Eddington-Born-Infeld action (EBI) for these degrees of freedom was proposed in
\cite{banados2}  and it was shown that they had unexpected effects: they could mimic the presence of dark energy and
dark matter in the expansion of the Universe and could modify the Newton-Poisson equation, leading to flat rotation
curves for galaxies. Hence it was proposed that the EBI action was a candidate for non-particulate dark matter and dark-energy.
The Eddington action \cite{eddington} has also been considered in the context of dark energy\cite{Gibbons}.

In this paper we wish to study the effect of EBI degrees of freedom on the expansion of the Universe and on the growth
of structure of the Universe. In doing so, we will identify the different regimes in the expansion rate and how they depend on
the parameters in the action and we will calculate the effect on the density perturbations and the cosmic microwave background (CMB). We will
focus on two possible uses for the EBI theory, one in which the extra degrees of freedom unify the dark sector, as proposed in \cite{banados1} and another
in which they co-exist with a cosmological constant, playing the role of dark matter. As a result we can identify a viable theory of
dark matter which is competitive with the standard CDM paradigm.

The paper is structured as follows. In the Section \ref{action} we display the EBI action and equations of motion and rewrite them as a specific case of {\it bigravity} or, alternatively, as a particular {\it bimetric} theory, as recently shown in \cite{ShortNote};
in Section \ref{cosmos} we study the dynamics of homogeneous and
isotropic solutions to the equations of motion (see~\cite{Davi} for homogeneous but anisotropic solutions);
in Section \ref{perturbations} we study the growth of linear perturbations and  in Section \ref{lss} we calculate
the power spectrum of density perturbations and anisotropies of the cosmic microwave background, allowing us to make
a cursory comparison with current data; finally in Section \ref{discussion} we discuss our findings.

\section{The theory: Eddington-Born-Infeld action as bi-gravity or as a bi-metric theory}
\label{action}
The EBI action is
\begin{eqnarray}\label{EBI}
 I &=& \frac{1}{16\pi G}\int d^4x \left[ \sqrt{-g}( R-2\Lambda)  + \frac{2}{\alpha \ell^2} \sqrt{ | \mathbf{g} - \ell^2 \mathbf{K} | }\right] \nonumber \\
  && \ \   + \ \  S_m[g]
\end{eqnarray}
where $\alpha$ is a dimensionless constant, $\ell$ a scale, $G$ is Newton's constant, $R$ is the scalar curvature of $g_{\mu\nu}$, and $S_m$ is the matter action.
The tensor $K_{\mu\nu}$ is the Ricci curvature of a connection $C^\alpha_{\mu\nu}$ defined in the usual way as
\begin{equation}
 K_{\mu\nu} = \partial_\alpha C^\alpha_{\mu\nu} - \partial_\nu C^\alpha_{\mu\alpha} + C^\alpha_{\alpha\beta} C^\beta_{\mu\nu} - C^\alpha_{\beta\mu}C^\beta_{\alpha\nu} \nonumber
\end{equation}
The connection $C^\alpha_{\mu\nu}$ should not be confused with the Christoffel connection $\Gamma^\alpha_{\mu\nu}$ of the metric $g_{\mu\nu}$. Note that in the limit in which $g\rightarrow 0$, the action for $q_{\mu\nu}$ reduces to the Eddington action \cite{eddington}.  The action (\ref{EBI}) is a functional of $g_{\mu\nu}$ and $C^{\alpha}_{\ \mu\nu}$ and is varied with respect to these fields.

It turns out that there is simpler formulation for this theory \cite{ShortNote}. Define a 2nd cosmological constant
\begin{equation}
\lambda \equiv \frac{\alpha}{\ell^2}
\end{equation}
and consider the action for two metrics $g_{\mu\nu}$ and $q_{\mu\nu}$
\begin{eqnarray}\label{biG}
 S &= &\frac{1}{16\pi G} \int d^4x [\sqrt{-g} (R-2\Lambda)  +  \sqrt{-q} ( K - 2\lambda) \nonumber \\
 & &-\sqrt{-q}\frac{1}{\ell^2} \invq^{\mu\nu} g_{\mu\nu} ]. \label{bigravity}
\end{eqnarray}
As shown  in \cite{ShortNote} the action (\ref{EBI}) is fully equivalent to (\ref{biG}).  The connection $C^{\mu}_{\ \alpha\beta}$ is related to the metric $q_{\mu\nu}$ by the usual metricity relation,
\begin{equation}
C^\alpha_{\mu\nu}  = \frac{1}{2}  \invq^{\alpha\beta}\left( \partial_\mu q_{\nu\beta} + \partial_\nu q_{\mu\beta} - \partial_\beta q_{\mu\nu}\right).
\nonumber
\end{equation}
where $\invq^{\mu\nu}$ is the inverse of $q_{\mu\nu}$ such that
\begin{equation}
\invq^{\mu\alpha} q_{\alpha\nu} = \delta^\mu_{\;\;\nu} \nonumber
\end{equation}
and $K \equiv (q^{-1})^{\mu\nu} K_{\mu\nu}$.

Theories of bigravity have been proposed in a number of contexts: as spin-2 theory of the strong interaction \cite{ISS,IshamStorey},
 as a full non-linear extension
of the Fierz-Pauli theory of massive gravity, and more recently as an effective theory of interacting brane-worlds \cite{DamourKogan}.    A number of examples
of bigravity theories have been studied in detail~\cite{DKP,AHGS,BDG,Deffayet},
in particular in terms of their consistency, asymptotic behavior and the global dynamics
of isotropic and homogeneous space times.
It should be remarked that in the context of bigravity theories, EBI turns out to be the simplest theory with a minimal interaction between both sectors.
 In this paper we  study in detail the cosmological dynamics both of the background and at the perturbative level and hence extract useful hints of what
 one might expect from more general classes of bigravity theories.
A class of bi-measure theories have been considered in~\cite{Guendelman} and references therein.

There is yet another point of view one can take of this theory. If one looks at the action as it is presented in equation \ref{bigravity}, i.e. a theory of two metrics, one of them, $g_{\mu\nu}$, quite clearly couples to the matter fields and has physical significance- it is this metric that defines how clocks and rulers respond-
and hence we can call it a ``physical" metric. The other metric, $q_{\mu\nu}$ satisfies the Einstein-Hilbert action and couples to the rest of the world through it's interaction with the physical metric. If we interpret $q_{\mu\nu}$ to be the metric of space time- we can dub it the ``geometric" metric- we then have a
bona-fide bimetric theory of gravity. This is entirely akin to the approach in the Tensor-Scalar-Vector theory of gravity \cite{teves} and can give us
an intriguing interpretation of roles of the different fields.

The field equations which are found from either the original EBI or the bigravity (or bimetric) action are
the Einstein equations for $g_{\mu\nu}$
\begin{equation}
 G^\mu_{\;\;\nu} = 8\pi GT^\mu_{\;\;\nu} - \Lambda \delta^{\mu}_{\;\;\nu}- \frac{1}{\ell^2} \sqrt{\frac{q}{g}} \; \invq^{\mu\alpha} g_{\alpha\nu}
\end{equation}
and the Einstein equations for $q_{\mu\nu}$
\begin{equation}
 Q^\mu_{\;\;\nu} =
- \lambda \delta^{\mu}_{\;\;\nu}
+ \frac{1}{\ell^2} \left[ (q^{-1})^{\mu\alpha} g_{\alpha\nu}
 - \frac{1}{2} (q^{-1})^{\alpha\beta} g_{\alpha\beta} \delta^\mu_{\;\;\nu}
\right]     \label{curv_EBI_eq}
\end{equation}
where $Q^\mu_{\;\;\nu} = K^{\mu}_{\;\;\nu} - \frac{1}{2} K \delta^{\mu}_{\;\;\nu}$ is the Einstein tensor of $q_{\mu\nu}$.
These are the complete set of equations with which we can study the dynamics of the EBI action. (Note that tracing (\ref{curv_EBI_eq}) a simplified equation is obtained $K_{\mu\nu}=\lambda q_{\mu\nu} + {1 \over \ell^2}g_{\mu\nu}.)$

\section{Cosmological dynamics}

\subsection{FLRW equations}
\label{cosmos}
We now focus on the dynamics of homogeneous and isotropic metrics in EBI gravity and will restrict ourselves to spatially flat metrics
so that the line element is given by
\begin{equation}
 ds^2 = -dt^2 + a^2 \gamma_{ij} dx^i dx^j
\nonumber
\end{equation}
where $t$ is physical time, ${\bf x}$ are spatial coordinates and $\gamma_{ij}$ is the metric of a flat hypersurface. Note that from
the bimetric point of view, this makes sense- all observables will depend on the physical metric, $g_{\mu\nu}$.
The vanishing of the Lie derivative for all Killing vectors of the spacetime gives $q_{\mu\nu}$ such that,
\begin{equation}\label{XY}
q_{00} = -X^2, \ \ \ \ \  q_{ij} = Y^2 \gamma_{ij}.
\end{equation}
The functions $X,Y$ parameterize the metrics compatible with the background symmetries.

The Friedmann equation for this cosmology is
\begin{equation}
3H^2 =   8\pi G  (\rhob_\E + \rhob_f )
\label{H_c}
\end{equation}
where $H = \frac{\dot{a}}{a}$, the EBI density is given by
\begin{equation}
\rhob_\E= \frac{Y^3}{8\pi G \ell^2 Xa^3}
\label{rho_E}
\end{equation}
which is always positive, and $\rhob_f$ is the energy density in all the remaining fluids (including the
cosmological constant, $\Lambda$).
The Raychaudhuri equation becomes
\begin{equation}
 - 2 \frac{ \ddot{a}}{a} - H^2  = 8\pi G  (\Pb_\E +  \Pb_f )
 \nonumber
\end{equation}
where $ \Pb_\E \equiv - XY/8\pi G \ell^2 a $ and $\Pb_f$ is the pressure in  all the remaining fluids (including the
cosmological constant, $\Lambda$).
The remaining field equations are then
\begin{eqnarray}
6 \ell^2 \frac{\dot{Y}^2}{Y^2} &=& 2 \alpha X^2    +  \frac{3a^2X^2}{Y^2}  -  1 \nonumber\\
  3\ell^2 \left[ \frac{\ddot{Y}}{Y} -  \frac{\dot{Y}}{Y} \frac{\dot{X}}{X}\right]  & =&  1+  \alpha X^2
\label{Y_dot}
 \end{eqnarray}

The EBI degrees of freedom behave as a fluid. We can trade the variables $X$ and $Y$ in terms of the fluid density $\rhob_\E$ and equation
of state parameter $w_\E$ given by
\begin{equation}
 w_\E =   - \frac{ a^2 X^2}{Y^2}  .
\label{def_wE}
\end{equation}
Eliminating the coordinate time $t$ using the Friedman equation (\ref{H_c}), we find that the new variables evolve as a function of $\ln(a)$ as
\begin{equation}
\rhob_\E' = - 3  (1 + w_\E) \rhob_\E
\label{rho_E_p}
\end{equation}
and
\begin{eqnarray}
  w_\E' =   2 w_\E  \bigg[  1 &+&  3w_\E
   \nonumber \\
&+&   \sqrt{ 4(-w_\E)^{3/2}\Omega_\E   \alpha - 2\frac{(1+ 3w_\E)\rho_\Ll}{\rho_c}} \bigg]
   \nonumber \\
\label{w_E_p}
\end{eqnarray}
where the relative density of a species "$i$" (including EBI) is as usual $\Omega_i = \frac{\rhob_i}{\rho_c}$ for the critical density
$\rho_c = \rhob_\E + \rhob_f$, and where we have defined
$
\rho_\Ll \equiv [8\pi G\ell^2]^{-1}
$.
Clearly the above equation is inconsistent for $w_\E>0$;  in fact $w_\E$ is bounded from above by the
condition
\begin{equation}
 16\pi G \ell^2  (-w_\E)^{3/2}\rho_\E   \alpha -3w_\E \ge  1.
\label{w_ineq}
\end{equation}

\subsection{Analytical approximations}

The system  (\ref{H_c}), (\ref{rho_E_p}) and (\ref{w_E_p}) can be solved numerically but we can extract some analytical results by examining its approximate behaviour.

The first clear case that one can see, is that when $w_\E \approx 0$, the equation of state $w_\E$ will evolve slowly and the
EBI field will behave as cold dark matter. This is independent of whether the EBI field is dominating the background dynamics or not,
and is therefore valid throughout the entire history of the universe (e.g. during radiation, matter and possible cosmological constant eras)
provided $|w_\E|$ is small. In this case $w_\E \approx w_0 a^2$ where
$w_0 <0$ is an initial condition. Since $w_\E$ is proportional to $a^2$,  the cold dark matter behaviour is unstable,
 and is bound to end when $w_\E$ is sufficiently driven away from zero and becomes $O(1)$. At this point, the subsequent behaviour of the
EBI field, depends on various factors which we analyse below on a case by case basis.

\subsubsection{Case : $\rho_\E \gg \rho_f$}
Let us first consider the case where $\rho_\E \gg \rho_f$ i.e the EBI field is driving the background dynamics.
Apart from the dark matter behaviour ($w_\E \approx 0 $) which as we have discussed above can  always be realized,
we uncover two more phases.
The first is a constant-$w$  phase such that $w_\E = w_{c}$  which  solves the equation
\begin{equation}
  (1+ 3w_c)^4 +  16\alpha^2 w_c^3   = 0
\end{equation}
This constant $w$ depends only on the parameter $\alpha$. For $\alpha = 0$ we have that $w_c = -\frac{1}{3}$ while for
$\alpha = 1$ we have that $w_c = -1$. This means that $-1<w_c< -\frac{1}{3}$ for $0<\alpha<1$ and $w_c <-1$ for $\alpha >1$. Indeed
in this case phantom behaviour is possible by allowing $\alpha>1$.


When $\alpha<1$ (i.e. $-1<w_c< -\frac{1}{3}$), the above constant-$w$ phase is unstable, simply because $\rho_\E$ eventually drops and approaches $\rho_\Ll$.
If this happens before the fluid $\rho_f$ becomes dominant (which could happen if $\rho_f$ is a cosmological constant),
then the constant-$w$ phase ends and the EBI fluid now behaves like a cosmological constant, i.e. $w_\E = -1$.
Eq. (\ref{w_E_p}), then gives that this cosmological constant is given by
\begin{equation}
 \rho_\E = \frac{\rho_\Ll}{1-\alpha}.
\end{equation}


It turns out that this deSitter phase is stable under  homogeneous time-dependent perturbations (although not under inhomogeneous perturbations;
  see relevant section below).
We postpone the stability analysis until case 3.

\subsubsection{Case : $\rhob_\E = \beta \rhob_f$ and $\rhob_f\ne \rho_\Lambda$ (tracking phase)}
An interesting case emerges if $\rhob_\E$ is neither negligible nor dominant but rather is assumed to track the
fluid. This is possible provided  $\rhob_\E=\beta\rhob_f$ where $\beta$ is a proportionality constant related to the
EBI relative density as $\Omega_\E = \beta/(1+\beta)$.
Eq (\ref{w_E_p}) then tells us  that this is possible iff $w < - \frac{1}{3}$, in which case we get
\begin{eqnarray}
\beta=\frac{(1+3w)^2}{4(-w)^{3/2}\alpha-(1+3w)^2} \nonumber
\end{eqnarray}
which gives $\Omega_\E = \frac{(1+3w)^2}{4(-w)^{3/2}\alpha}$.

Since there are no interesting fluids with equation of state $w<-1/3$ (apart from a cosmological constant, treated below), this
case is not of much relevance.

\subsubsection{Case : $\rhob_\E = \beta \rho_\Lambda$  (cosmological constant tracking phase)}
In the limit where $w=-1$ we find $\beta=1/(\alpha-1)$ and therefore  $\rho_E=\rho_\Lambda/(\alpha-1)$,  which is valid only for $\alpha>1$.

We can explore this limit further; if $\rho_E$ is constant and $\rho_\Ll$ non-negligible then
we have that the EBI density is given by
\begin{equation}
\rho_\E = \frac{\rho_\Ll - \rho_\Lambda}{1-\alpha}
\end{equation}
while the effective cosmological constant  such that $3 H^2 = 8\pi G \rho^{(\text{eff})}_{\Lambda}$ is
\begin{equation}
\rho^{(\text{eff})}_{\Lambda}= \rho_\E + \rho_\Lambda =\frac{\rho_\Ll - \alpha \rho_\Lambda}{1-\alpha}
\end{equation}
In the limit in which $\ell\rightarrow\infty$ we recover the previous case.

One should further impose the conditions $\rho_\E \ge 0$ and $\rho^{(\text{eff})}_{\Lambda}>0$.
For $0<\alpha<1$, a necessary and sufficient condition for this to hold, is that $\rho_\Ll > \rho_\Lambda$ (regardless of the sign of $\rho_\Lambda$),
while for $\alpha>1$ we need  $0\le \rho_\Ll < \rho_\Lambda$. This second subcase cannot be realized (see case-5 below).
For $\alpha<1$, taking the limit $\rho_\Lambda \rightarrow 0$ takes us back to case-1.

The negative sign appearing in the expressions above is quite misleading, and one could think that
it might be possible to cancel the effective cosmological constant to sufficiently small values.
This is clearly impossible for $\rho_\Lambda>0$ simply because by virtue of (\ref{rho_E}) we also have  $\rho_\E>0$.
It is also impossible for  $\rho_\Lambda<0$ since again because of  (\ref{rho_E}) we need $\alpha<1$ which  implies
that $\rho^{(\text{eff})}_{\Lambda}> \rho_\E > |\rho_\Lambda| $.
 Thus we cannot have cancellation of the cosmological constant.

We now perform stability analysis (as mentioned in case-1) of this deSitter phase (for which $\alpha<1$).
Let $\rhob_\E = \frac{\rho_\Ll-\rho_\Lambda}{1-\alpha}(1 + \epsilon_1)$ and $w_\E = -1 + \epsilon_2$ with $\epsilon_1>0$ and $\epsilon_2>0$.
Perturbing (\ref{rho_E_p}) and (\ref{w_E_p})  to linear order we find
\begin{eqnarray}
{\epsilon_1}' = - 3 \epsilon_2 \nonumber
\end{eqnarray}
and
\begin{eqnarray}
 {\epsilon_2}' = 2(1-\alpha)\frac{\rho_\Ll - \rho_\Lambda}{\rho_\Ll - \alpha\rho_\Lambda} \epsilon_1 - 3 \epsilon_2 \nonumber
\end{eqnarray}
which combine to give
\begin{eqnarray}
{\epsilon_1}'' +  3 {\epsilon_1}' +  6(1-\alpha)\frac{\rho_\Ll - \rho_\Lambda}{\rho_\Ll - \alpha\rho_\Lambda} \epsilon_1 = 0 \nonumber
\end{eqnarray}
The normal modes are  $e^{n\ln a}$ where
\begin{equation}
n = \frac{3}{2}\left[ -1 \pm\sqrt{1 - \frac{8}{3}(1-\alpha)\frac{\rho_\Ll - \rho_\Lambda}{\rho_\Ll - \alpha\rho_\Lambda} } \;\;\right] \nonumber
\end{equation}
Therefore the approach to deSitter is critically damped for $\alpha = \alpha_c = \frac{5\rho_\Ll - 8\rho_\Lambda}{8\rho_\Ll - 11  \rho_\Lambda}$ while it is
underdamped for $\alpha<\alpha_c$ and overdamped for $\alpha>\alpha_c$.
Furthermore, we have that $-\frac{5}{3}< 1 - \frac{8}{3}(1-\alpha)\frac{\rho_\Ll - \rho_\Lambda}{\rho_\Ll - \alpha\rho_\Lambda} <1$ and so
the underdamped solutions are always decaying. Finally it is possible to have only the overdamped solutions
by choosing $\rho_\Lambda <\rho_\Ll < \frac{8}{5}\rho_\Lambda$.

\subsubsection{Case : $\rho_\E \ll \rho_f \ne \rho_\Lambda$ }
We now pass to the regime where $\rho_\E \ll \rho_f$, i.e. the EBI field is subdominant, and the cosmological
dynamics are driven by some fluid $\rho_f$ which is \emph{not} a cosmological constant. Consistency requires that $\rho_\Ll \ll \rho_f$ (otherwise
this case cannot be realized), and we therefore get that
\begin{eqnarray}
  w_\E' \approx   2 w_\E (1+ 3w_\E )
\label{w_E_case_1B}
\end{eqnarray}
Hence we find that if the EBI-field is subdominant, the above equations lead to two possible behaviours : the EBI fluid behaves either as cold dark matter if $0 < w_\E \ll -1$, i.e.
very close to zero, or as curvature if $w_\E \sim -\frac{1}{3}$.

\subsubsection{Case : $\rho_\E \ll \rho_f = \rho_\Lambda$ }
The final case we consider is when $\rho_\E$ is negligible but now $\rho_f = \rho_\Lambda$,
i.e. the background fluid which drives the dynamics is the bare cosmological constant $\Lambda$. Here
 we find a new regime such that
\begin{eqnarray}
w_\E = w_\Ll \equiv -\frac{1}{3}-\frac{2}{3}\frac{\rho_\Ll}{\rho_\Lambda}
\end{eqnarray}
 in addition to the cold dark matter regime which can still be realized.
Note that we recover the curvature like behaviour if $\rho_\Ll\ll \rho_\Lambda$ while $w_\E=-1$ if $\rho_\Lambda=\rho_\Ll = (8\pi G \ell^2)^{-1}$.

One may wonder whether phantom behaviour such that $w_\E <-1$ can be realized in this case, by choosing $\ell$  such that $ 8\pi G \ell^2 \rho_\Lambda< 1$.
This turns out to be impossible: when $\rho_\Lambda$ becomes smaller than the threshold value  $8\pi G \ell^2 \rho_\Lambda=1$, this takes us back
to case 3. Decreasing $\ell$ further (or decreasing $\rho_\Lambda$) eventually leads to case 1.

\subsection{Realistic model building for the background dynamics}
Having analyzed the different possible behaviours of the EBI field in various cases above, we now turn
to realistic model building.

We will use units in $Mpc$ which is the standard in popular Boltzmann solvers such as
CMBfast~\cite{cmbfast}, CAMB~\cite{camb} and CMBeasy~\cite{cmbeasy}. In these units we have that
 the Hubble constant today is $H_0 = 3.34 \times 10^{-4} \; h \; Mpc^{-1}$, with $h\sim 0.6-0.8$.
For numerics we can absorb $8\pi G $ into the definition of densities. The total fluid density is thus
\begin{equation}
8\pi G \rhob_f =  3.34 \times 10^{-7}  \left[ \frac{\omega_r}{a^4} + \frac{\omega_b}{a^3} + \omega_\Lambda\right]  Mpc^{-2}
\end{equation}
where $\omega_r =  4.16\times 10^{-5}$ (for CMB temperature of $2.726K$ and three species of massless neutrinos),
 $\omega_b \sim 0.018 - 0.023$ given by nucleosynthesis, and $\omega_\Lambda \sim 0 - 0.5$.

Turning to the EBI field, we need to set its initial density $\rhob_{\E,\text{in}}$ and initial equation of state parameter $w_{\E,\text{in}}$
at the initial scale factor $a_i$. To do this we require that
 the initial condition for $\rhob_\E$ is such that it would give rise to an equivalent CDM density in the past. In other words
requiring that the equivalent CDM density today would be $8\pi G \rhob_c =  3.34 \times 10^{-7} \;\omega_c\; Mpc^{-2}$, with
$\omega_c \sim 0.11$, we extrapolate this to the initial scale factor $a_i$ and set the initial condition for the
EBI density as $8\pi G \rhob_{\E,\text{in}} =   3.34 \times 10^{-7}\times a_i^{-3} \;\omega_\E \; Mpc^{-2}  $,
with $\omega_\E \sim 0.08 - 0.13$.

For setting the initial condition  for $w_\E$, we require that the EBI field behaves as CDM all the way up to at least $a\sim 0.1$ where
$w_\E$ starts to become $O(1)$. Since in the CDM phase $w_\E \sim  -w_0 a^2$, we set the initial condition for $w_\E$
as $w_{\E,in} = -w_0 a_i^2$, by specifying a positive parameter $w_0$.
\begin{figure}
\epsfig{file=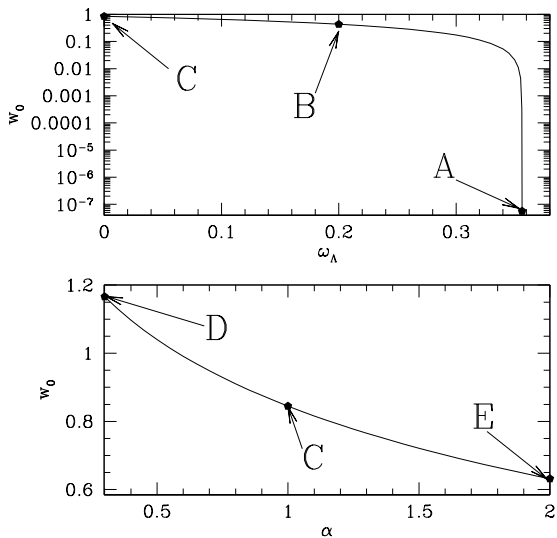,width=3.in}
\caption{Upper panel : The locus of fixed $\tau_0$ and angular diameter distance to recombination, by varying $\omega_L$ and $w_0$ keeping all
other parameters fixed. The value $\omega_\Lambda = 0.36$ corresponds to the WMAP5 best fit model. The other EBI parameters are $1-\alpha = 10^{-6}$
 and $\ell = 10^9 Mpc$.  Lower panel : A similar locus curve, only now we vary $\alpha$ and $w_0$ keeping all other
parameters fixed and in particular $\omega_\Lambda = 0$.
 The alphabetical labels are explained and discussed in the main text. }
\label{equal_dist}
\end{figure}

The background model is thus completely determined by six parameters : the initial conditions
 $\omega_b$, $\omega_\Lambda$, $\omega_\E$ and $w_0$ as well as the two parameters $\ell$ and $\alpha$ (for the
fixed radiation density discussed above; massive neutrinos can easily be accommodated in the usual way but we refrain to discuss it here
for reasons of simplicity).
On top of specifying these parameters one has to make sure that the inequality constraint (\ref{w_ineq}) is obeyed. In the light of
setting up initial conditions as we have just described the inequality becomes
\begin{equation}
\alpha  \left(\frac{\ell}{Mpc}\right)^2   w_0^{3/2} \omega_\E \ge 1.5  \times 10^6.
\end{equation}

\subsubsection{The $\Lambda$EBI model}
The simplest possibility is when the EBI field is chosen to act as CDM all the way, and is not responsible for the
accelerated expansion of the universe which is due to the bare cosmological constant $\Lambda$. In this case,
in order for the EBI field not to deviate from the CDM track, we must have that $w_0 \le 10^{-4}$. Moreover
the parameters $\ell$ and $\alpha$ do not have any role in the background.
 Thus we are down to three parameters :  $\omega_b$, $\omega_\Lambda$ and
 $\omega_\E$, the same as in the standard $\Lambda$CDM model. We call this model the $\Lambda$EBI model.

Choosing $\omega_b = 0.023$, $\omega_\Lambda = 0.36$ and $\omega_\E = 0.114$ we have a background evolution identical
to the WMAP5 best fit $\Lambda$CDM model. It is therefore not distinguishable from the $\Lambda$CDM model using, for example,
type-1a supernovae data~\cite{SN1a}.

\subsubsection{The general EBI model}
By lowering the cosmological constant $w_\Lambda$ gradually to zero, we should start compensating by having the EBI field to play a
role in the accelerated expansion of the universe.
\begin{figure}
\epsfig{file=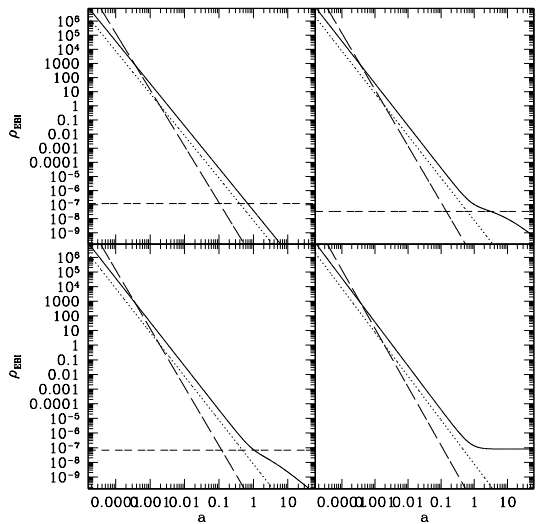,width=3.in}
\caption{Four models from the locus curve of Fig.\ref{equal_dist} $(\omega_\Lambda,w_0)$ =  $(0.36,6\times10^{-8})$ (upper left),
 $(0.2,0.43)$ (upper right), $(0.1,0.65)$ (lower left), $(0.,0.84)$ (lower right). The curves are : EBI (solid), $\Lambda$ (dashed),
baryons (dotted) and radiation (long-dashed). Note that in the last case (lower right),  the final state of the EBI field is
an approximate cosmological constant. The actual phase is the constant-w phase with w being extremely close to $-1$ due to $1 -\alpha = 10^{-6}$.
}
\label{fig_rhoE_real_var}
\end{figure}

\begin{figure}
\epsfig{file=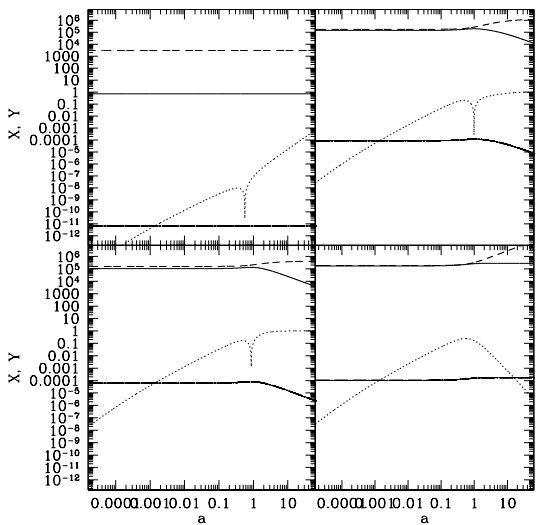,width=3.in}
\caption{Same four models as Fig.\ref{fig_rhoE_real_var}.
  They are from the locus curve of Fig.\ref{equal_dist} $(\omega_\Lambda,w_0)$ =  $(0.36,6\times10^{-8})$ (upper left),
 $(0.2,0.43)$ (upper right), $(0.1,0.65)$ (lower left), $(0.,0.84)$ (lower right). The curves are : $X$ (solid), $Y$ (dashed),
$\frac{\dot{X}}{X} $(dotted) and $\frac{\dot{Y}}{Y}$ (long-dashed).
}
\label{fig_XY_real_var}
\end{figure}

Lowering $\Lambda$ changes the angular diameter distance to the surface of last scatter and hence  shifts the position of the acoustic peaks in
the CMB. Since the peaks
are very tightly constrained, we must change one further parameter to make up for the change induced by varying $\Lambda$.
As a 2nd parameter (for illustration) we choose to vary $w_0$. The upper panel of figure \ref{equal_dist} shows the
locus curve for which both the angular diameter distance and conformal time today $\tau_0$, remain constant, in the $w_0$ and $\omega_\Lambda$ plane.
The point $\omega_\Lambda =  0.36$ and $w_0 = 6\times 10^{-8}$. We call this model "Model A". The
parameters $\alpha$ and $\ell$ are chosen to be $|1 - \alpha| = 10^{-6}$ and $\ell = 10^{9}Mpc$ for this model. Increasing $\ell$ or lowering $w_0$
still gives acceptable $\Lambda$EBI models as the dependence on lowering $w_0$ or increasing $\ell$ is very week and for most cases
does not produce any observable result. As we lower $\omega_\Lambda$ towards zero, we keep increasing $w_0$, along the displayed
curve to make sure that the angular diameter distance to recombination stays constant.
 When $\omega_\Lambda =0$ exactly, $w_0 \approx 0.845$ (this depends on our choice of $\omega_b$ and $\omega_\E$).
We call this "Model C". For clarity we also consider a "Model B" for which $\omega_\Lambda = 0.2$ and $w_0 \approx 0.43$.
Figure \ref{fig_rhoE_real_var} shows the
evolution of the EBI energy density (solid), radiation density (long dash), baryons (dotted) and cosmological constant (dash) for four models
along this curve. The upper left panel is Model A, the upper right panel is  model B with $\omega_\Lambda = 0.2$ and $w_0\approx 0.43$,
the lower right panel is a model with $\omega_\Lambda = 0.1$ and $w_0\approx 0.65$, and finally the lower right panel is model C.
Figure \ref{fig_XY_real_var} exhibits the variables $X$ (solid), $Y$ (dashed), $\frac{\dot{X}}{X}$ (dotted) and $\frac{\dot{Y}}{Y}$ (long--dashed)
for the same set of models. Observe that during the time for which the EBI field is like CDM, both $X$ and $Y$ are approximately constant (very
slowly varying), while during the $w_\E\approx -1$ phase, $X$ is still an approximate constant while $Y$ is varying.

Once we reach the point $\omega_\Lambda = 0$ (model C) we can start investigating the effect of changing $\alpha$. Once again we
keep changing $w_0$ in order to compensate and keep the angular diameter distance the same.
Along this line we consider two further models :
Model D has $\alpha=0.3$ and $w_0\approx 1.166$, while Model E has $\alpha = 2$ and $w_0 \approx 0.632$.
Remember that $\alpha$ is connected with the constant $w_\E$ phace. In particular Model D has $w_\E \approx -0.493$ during the acceleration era
while Model E gives rise to phantom behaviour with $w_\E \approx -1.795$.

In principle we can further investigate varying $\ell$. We find however that changing $\ell$ does not lead to any interesting new behaviour.
If we compensate the variation of $\ell$ with $w_0$, then $w_0$ must be increased, and this leads to a similar effect as decreasing $\alpha$.

Let us also note that the general EBI family of models would give a background evolution that deviates from the $\Lambda$CDM model.
They can therefore be distinguished from $\Lambda$CDM using, for example, type-1a supernovae data~\cite{SN1a}. However, as we discuss further below,
we do not find any parameter space allowed (apart from the $\Lambda$EBI special case) when we consider the Cosmic Microwave Background
angular power spectrum observations. It is therefore of little significance to try to constrain such models with the supernovae data.

\subsection{Summary of the background evolution}
In this section we have mapped out various possibilities for the background evolution.
 As claimed in \cite{banados2}, it is possible to construct a theory in which the EBI field plays the dual role of both dark matter and dark energy.
We have generalized the results in  \cite{banados2} by uncovering a third phase of the EBI field where it has a constant-w equation of state which
interpolates between the CDM phase and the cosmological constant phase. Thus the EBI field is a unified model very similar to
 to Chaplygin gas. We stress however that contrary to the Chaplygin gas,  $w_\E$ is an independent dynamical degree of freedom, and thus
the equation of state of EBI is not rigid. This leads to even richer  dynamics in the perturbations as we show in the next section.

We have specified the requirements on the initial conditions of the EBI field as well as the two parameters $\ell$ and $\alpha$, in order
to have a background evolution that is compatible with the standard paradigm. This gave us the simple  $\Lambda$EBI model, for which the EBI
field replaces CDM but is not responsible for the accelerated expansion which is still due to the cosmological constant $\Lambda$. By gradually
lowering $\Lambda$ to zero, and compensating with the EBI field, one can have the initial condition $w_0$ as well as the parameters $\ell$ and $\alpha$ to
play a role, leading to the mixed EBI model, where the effective cosmological constant recieves a contribution from $\Lambda$ and the EBI field. When
$\Lambda =0$ we get the plain EBI model.

\section{Evolution of inhomogeneities}
\label{perturbations}

As we have seen in previous sections, the EBI background field can behave as pressureless matter and as cosmological constant, and there are transitions between these two phases.

After displaying the equations of motion for linearized fluctuations, in this section we show via analytical approximations that the matter phase is consistent with current observations. On the contrary, fluctuations on the acceleration phase show an unacceptable growth.  Finally, we show an analytical series in $1/\ell$ providing a systematic way to isolate the matter phase, as the zero order approximation, leaving an evolution indistinguishable from $\Lambda$CDM.  Corrections in powers $1/\ell$ can then be computed to any desired order. These will be reported elsewhere.

\subsection{The equations of motion}

For the purpose of studying large scale structure, i.e. density perturbations, we will focus on scalar perturbations in the conformal
Newtonian gauge in conformal time, $\tau$, such
that $g_{00} = -a^2(1 + 2 \Psi)$, $g_{0i} = 0$, $g_{ij} = a^2(1 - 2\Phi) \gamma_{ij}$. We have that
$\grad_i$ is the covariant derivative on the hypersurface such that $\grad_i \gamma_{jk} = 0$,  and define
 $\TD_{ij} = \grad_i \grad_j - \frac{1}{3} \gamma_{ij} \grad^2$.

The tensor field $q_{\mu\nu}$ is perturbed as $q_{00} = -a^2X^2(1 + 2 \Xi)$, $q_{0i} = -Y^2 \grad_i \beta$, $q_{ij} =  Y^2\left[(1 - 2\chi) \gamma_{ij}
 +  \TD_{ij} \mu \right]$. Notice that the $q$-metric has four scalar modes, namely $\Xi$, $\beta$, $\chi$ and $\nu$, as there is no
gauge freedom left to set any of them to zero.
We also find it convenient to define $Z=  \frac{d\ln Y}{d\tau}$

As in the homogeneous case, it turns out that the EBI field can be cast as a generalized fluid, in the
framework of a generalized dark matter model~\cite{Hu}.
The Einstein equations are
\begin{eqnarray}
-2 k^2  \Phi -6 \frac{\dot{a}}{a} \dot{\Phi} - 6 \frac{\dot{a}^2}{a^2} \Psi &=&  8\pi G a^2 \sum_i \rhob_i \delta_i
\\
2\dot{\Phi} + 2 \frac{\dot{a}}{a}  \Psi &=& 8\pi G a^2  \sum_i \rhob_i  \Theta_i
\\
6\ddot{\Phi}
+ 6\frac{\dot{a}}{a} \left[2\dot{\Phi}  +\dot{\Psi} \right]
 + 2k^2 (\Phi &-& \Psi)
  \nonumber \\
 - 6 \left[-2 \frac{\ddot{a}}{a} + \frac{\dot{a}^2}{a^2}\right] \Psi
 &=& 24 \pi G a^2  \sum_i \rhob_i \Pi_i
\\
 \Phi -\Psi &=& 8\pi G a^2  \sum_i \rhob_i  S_i
\label{PhiPsi}
\end{eqnarray}
where the index $"i"$, runs over all fluids, including the EBI field. Here $\delta$ is the fluid density fluctuation,
$\Theta$ is the fluid momentum divergence, $\Pi$ is the pressure fluctuation and $S$ is the fluid shear.
We can combine them  to find a Newton-Poisson-like equation of the form:
\begin{eqnarray}
-2 k^2  \Phi  &=& 8\pi G a^2  \sum_i \left[\rhob_i \delta_i  +3\frac{\dot{a}}{a} \rhob_i \Theta_i\right] \label{NP}
\end{eqnarray}

The EBI density contrast $\delta_\E$, velocity perturbation $\Theta_\E$, relative pressure perturbation $\Pi_\E$ and
shear perturbation, $S_\E$,  are defined as linear combinations of the EBI metric variables $\Xi$, $\beta$, $\mu$ and $\chi$ :
\begin{eqnarray}
 \delta_\E &=&  \Psi - \Xi + 3 (\Phi - \chi)
\\
 \Theta_\E &=&   -\beta
\\
 \Pi_\E &=&  w_\E ( \Xi - \Psi + \Phi - \chi)
\\
 S &=&   - w_\E\mu  \label{fluid}
\end{eqnarray}

The evolution equations for the above fluid variables are found to be
\begin{eqnarray}
\dot{\delta}_\E  &= & -  k^2\Theta_\E + 3(1 + w_\E)  \dot{\Phi} +3\frac{\dot{a}}{a} \left(w_\E\delta_\E - \Pi_\E \right)  \\
 \dot{\Theta}_\E & =&  \frac{\dot{a}}{a} ( 3w_\E-1) \Theta_\E +  (1 + w_\E)\Psi  -\frac{2}{3}  k^2 S_\E + \Pi_\E  \\
\dot{S}_\E &=& \left[ 4Z +2(1+3w_\E)\frac{\dot{a}}{a}  - \frac{w_\E k^2}{3Z}  \right]S_\E
\nonumber \\
 &-&2 w_\E \left[1 + \frac{3a^2}{2\ell^2 k^2}\right]  \Theta -2\frac{w_\E^2}{Z} \Phi
\nonumber \\
&+& \frac{w_\E}{2Z}\left[w_\E  + \frac{a^2 (3 w_\E -1 )}{2\ell^2 k^2} \right] \delta_\E
\nonumber \\
& + &\frac{1}{2Z}\left[w_\E + \frac{3a^2(1+w_\E)}{2\ell^2 k^2} \right] \Pi_\E  \\
\dot{\Pi}_\E &=&
  \left[7Z
+ \frac{a^2(1 + w_\E)}{2\ell^2Z}
+ \frac{k^2}{3Z} w_\E
 + (2+9w_\E)\frac{\dot{a}}{a} \right]\Pi_\E \nonumber \\ & &
+ w_\E \left[
- Z
-3\frac{\dot{a}}{a} w_\E
+ \frac{(3w_\E-1)a^2}{6\ell^2Z}
+ \frac{k^2}{3Z}  w_\E
\right]\delta_\E
\nonumber
\\
&&
-\frac{1}{3}k^2 w_\E \left[ \Theta + \frac{2k^2}{3Z}  S_\E \right] \nonumber \\ & &
+ w_\E \left[
4Z \Psi
- \frac{4k^2}{3Z}  w_\E \Phi
+ (1-3w_\E) \dot{\Phi}
\right]
\end{eqnarray}
The remaining fluids can be described in the usual way using conservation of energy and momentum.

\subsection{EBI as CDM}
\label{ebicdm}
In this section we show how the EBI field can behave as CDM even at the perturbative level.
First notice that after setting $\Pi_\E$ and $S_\E$ to zero in all of the fluctuation equations, the remaining two variables,
namely $\delta_\E$ and $\Theta_\E$ will obey equations which are the same as for a CDM fluid, provided
the background equation of state parameter $w_\E$ is also very close to zero and the EBI field is in the
background CDM phase.

We now show that if $w_\E\approx 0$ and the pressure perturbation $\Pi_\E$ and shear $S_\E$ are initially chosen to
be zero, then they will remain arbitrarily small, and the EBI field will behave as CDM even at the fluctuation level.

During this phase we get that $Z \approx  a H \sqrt{\alpha \Omega_\E }  (-w_\E)^{3/4} $, hence we can set
 $w_\E \rightarrow 0$, $Z \rightarrow 0 $ and $w_\E/Z \rightarrow 0$.
The $\Pi_\E$ equation then becomes
\begin{equation}
 \frac{d\Pi_\E}{d\ln a} = \left[ \frac{1}{2\ell^2 H^2  \sqrt{\alpha \Omega_\E }  (-w_\E)^{3/4} } + 2 \right]\Pi_\E
\end{equation}
Since it is not sourced by any other variable in this limit, if we set $\Pi_\E = 0$ initially, it will stay zero.
Thus imposing $\Pi_\E = 0$ in the shear equation and taking the same limit above we get
\begin{equation}
 \frac{dS_\E}{d\ln a} = 2  S_\E
\end{equation}
which has solution $S_\E = S_{0} a^2$.

Thus for  very small initial relative pressure perturbation and shear, i.e. $\Pi_0\simeq0$ and $S_\E \simeq 0$,
 the EBI field will behave exactly as cold dark matter at the perturbative level,
i.e. the equations for $\delta_\E$ and $\Theta_\E$ would read
\begin{eqnarray}
\dot{\delta}_\E  &= & -  k^2\Theta_\E + 3  \dot{\Phi}    \\
 \dot{\Theta}_\E & =& - \frac{\dot{a}}{a}  \Theta_\E +  \Psi   \label{matterf}
\end{eqnarray}
respectively.

This means that if $|w_\E|\ll1$ throughout the entire history of the universe until today (such as the $\Lambda$EBI model-A),
we would expect any observable to be  completely indistinguishable between an EBI dark matter model and a standard dark matter model.
Note also that we have assumed that $\Pi_0\simeq0$ and $S_\E \simeq 0$ to obtain {\it exact} CDM-like behaviour. But if
we were to include a small amount of pressure perturbation and shear in the initial conditions on very small
scales, this might lead to differences with the CDM which might be observable at the cores of galaxies, clusters and in
the small scale structure of gravitating bodies. Indeed, differences vis-a-vis CDM are expected to occur once
a system enters the non-linear regime where $\Pi_0\simeq0$ and $S_\E \simeq 0$ are inevitably sourced.

\subsection{EBI acceleration era}

The evolution of perturbations during a regime of accelerated expansion is more
intriguing.  Assume that $\Lambda=0$ and neglect baryons and other components. As shown in \cite{banados2} the equations of motion for EBI gravity have an exact de-Sitter solution,
\begin{equation}\label{dS}
a(\tau)= {\sqrt{3(1-\alpha)}\, \ell \over \tau_0-\tau}, \ \    X(\tau)={1 \over \sqrt{1-\alpha}}, \ \   Y(\tau) = {\sqrt{3}\, \ell \over \tau_0 - \tau} \nonumber
\end{equation}
This field solves equations (\ref{H_c}) and (\ref{Y_dot}).  $\tau_0$ is an arbitrary integration constant fixed by the initial conditions. $a\rightarrow \infty$ as $\tau\rightarrow\tau_0$.

To check whether this solution can represent or not the  accelerated era of the Universe we study fluctuations on this background. This is an straightforward exercise and we summarize here the main results. Using the equations of motion all EBI functions $\chi(\tau),\beta(\tau),\Xi(\tau),\mu(\tau)$ can be written as functions of the Newton potentials $\Phi(\tau)$ and $\Psi(\tau)$ by algebraic expressions.  We are left we two coupled second order differential equations for $\Phi(\tau)$ and $\Psi(\tau)$. Interestingly, the combination $\Phi(\tau)+\Psi(\tau)$ decouples and satisfy a Bessel-like equation with the exact solution,
\begin{equation}\label{}
\Phi(\tau)+\Psi(\tau) = {a_0 J_\nu(k(\tau_0-\tau)) + b_0 Y_\nu(k(\tau_0-\tau)) \over \sqrt{\tau_0-\tau}} \label{isw_pot} \nonumber
\end{equation}
where
\begin{equation}\label{}
\nu = {1 \over 2} \sqrt{24\alpha-39}.
\end{equation}
For any value of $\alpha$ this function diverges as $\tau\rightarrow \tau_0$ making de Sitter space unstable.

We would like to stress that this conclusion may change when introducing extra ``Fierz-Pauli" couplings to the EBI action.  See \cite{ShortNote} for a recent discussion.

\subsection{$1/\ell$ and the $\Lambda$\!EBI theory}

In this paragraph we develop an analytical method to isolate the matter phase for the EBI field. Numerical analysis show that this occur for $\ell$ and $Y$ large. We then attempt to use $1/\ell$, the coupling between both metrics, as a perturbative parameter.

Consider the following  Frobenious type series for the background functions  $a(t),X(t),Y(t)$,
\begin{eqnarray}
  a(\tau) &=& a_0(\tau) + {1 \over \ell} a_1(\tau) + {1 \over \ell^2} a_2(\tau) + \cdots\nonumber\\
  X(\tau) &=& X_0(\tau) + {1 \over \ell} X_1(\tau) + {1 \over \ell^2} X_2(\tau) + \cdots\nonumber\\
  Y(\tau) &=& \ell^{2/3}\left( Y_0(\tau) + {1 \over \ell} Y_1(\tau) + {1 \over \ell^2} Y_2(\tau) + \cdots\right) \label{1/l}
\end{eqnarray}
We have included a positive power of $\ell$ in $Y$ for two reasons. First, recall the EBI background density has the form $\rho_E \sim {1 \over \ell^2}{Y^3 \over Xa^3}$.  Thus, if $Y$ scales as $Y \sim \ell^{2/3}$, then at order zero in $1/\ell$ there will be a finite contribution to the Friedmann equation from the EBI field, which turns out to be dark matter.  The interesting observation is that this series also provides the right equations for all other variables including fluctuations.

There is another reason to include a growing $\ell$ term in the background.  If the metric functions do not depend on $\ell$, then in the limit of large $\ell$ the metrics $g_{\mu\nu}$ and $q_{\mu\nu}$ become decoupled.  This is clearly seen from the bi-gravity action (\ref{biG}).  The decoupled system describes two massless gravitons and is a theory with a different number of degrees of freedom. Switching on the interaction term, proportional to $1/\ell$, is in this sense a discontinuous change to the theory.  On the contrary, for a series of the form (\ref{1/l}) the two metrics never decouple.

We first discuss the background equations and then the fluctuations.  Plugging (\ref{1/l}) in the background equations we find consistent equations for the coefficients  $a_i(\tau),X_i(\tau),Y_i(\tau)$ order by order in $l$.

At order zero one finds that $X_0$ and $Y_0$ must be constants while $a_0(\tau)$ satisfies the Friedman equation
\begin{equation}\label{}
 {\dot a^2_0(\tau) \over a^4_0(\tau)} = {Y_0^3 \over 3X_0 }{1 \over a^3_0(\tau)} + {1 \over 3}\Lambda
\end{equation}
Thus, by choosing $Y_0^3/X_0$ and $\Lambda$ appropriately we find a background evolution which is, at this order, indistinguishable from $\Lambda$CDM.

Corrections $1/\ell$ can be computed order by order. Since $\ell$ is so far arbitrary we can make it as large as necessary in order to suppress $1/\ell$ corrections. For completeness we display the first order equations
\begin{eqnarray}
  3{dY_1 \over d\tau}X_0 - {dX_1 \over d\tau}Y_0 &=& 0 \\
  6 \left( {dY_1 \over d\tau} \right)^2 - 2\alpha X_0^2 Y_0^2 a_0(\tau) - Y_0^2 a_0(\tau)^2  &=& 0.
\end{eqnarray}
The equation for $a_1(\tau)$ is longer and not really worth displaying. The important point is that once $a_0(\tau)$ is known, the first order equations can be solved.

~

We now explore the fluctuation equations in the same limit assuming that the background satisfies $X(t)=X_0$, $Y(\tau) = Y_0 \ell^{2/3}$.  In this paragraph we work in terms of the original metric variables $\Psi,\Phi,\chi,\beta,\Xi,\chi$.   We only consider here the leading terms. One finds that the Newton potential $\Phi(\tau)$ satisfies the usual equation (recall that for matter $c^2_s=0$))
\begin{equation}\label{eP}
\ddot\Phi + 3 {\cal H} \dot \Phi  + \left(2 \dot  {\cal H} + {\cal H}^2 \right)\Phi =0
\end{equation}
where ${\cal H} = \dot a/a$. (We drop the subscript $_{0}$ in $a_{0}$ because we work only to this order and no confusion can arise.)  For matter $a = a_0 \tau^2$ and we recover, as already shown in section \ref{ebicdm}, the familiar $\Phi(\tau) = a_0 + b_0/\tau^5$, while for acceleration with $a \sim 1/(\tau_0-\tau)$ we have $\Phi = a_1 (\tau_0-\tau) + b_1 (\tau_0-\tau)^3$.

All other functions are expressed via the field equations in terms of $\Phi$ as follows:
\begin{eqnarray}
  \Psi(\tau) &=& \Phi(\tau) \nonumber\\
  \beta(\tau) &=& -{2X_0 \over Y_0^3} {d \over d\tau} (\Phi(\tau) a(\tau))  \nonumber\\
  \chi(\tau) &=& -k^2\mu(\tau) + c_1  \nonumber\\
  \Xi(\tau) &=& \Phi(\tau) + {2X_0 \over Y_0^3} \, k^2\Phi(\tau)a(\tau) - {1 \over 2}k^2\mu(\tau)  + c_2  \nonumber\\
 \mu(\tau)  &=&  {4X_0 \over Y_0^3} \Phi(\tau) a(\tau) + c_3 \int d\tau\, a(\tau)\label{fq}
\end{eqnarray}
where $c_1,c_2,c_3$ are integration constants.   Recalling the fluid variables (\ref{fluid}) it is direct to prove from here that both (\ref{matterf}) and the Newton-Poisson equation (\ref{NP}) are satisfied.

The $1/\ell$ series provides a systematic way to isolate the dark matter phase. At the same time, it provides a way to compute order by order deviations from $\Lambda$CDM which may reveal new features. We shall study these corrections elsewhere.

\section{The Cosmic Microwave Background and Large Scale Structure}
\label{lss}
The main goal in our analysis is to estimate the two main cosmological observables: the CMB
and the large scale structure of the distribution of galaxies in the Universe.

The anisotropies in the CMB can be described in terms of fluctuations in temperature, $(\Delta T/T)({\bf n})=[T({\bf n})-T_0]/T_0$, where
$T_0$ is the average temperature in the CMB and $T({\bf n})$ is the temperature measured in the direction ${\bf n}$. It is
convenient to look at the variance of these fluctuations expanded in Legendre Polynomials, $P_\ell$ such that
\begin{eqnarray}
\langle \frac{\Delta T}{T}({\bf n})\frac{\Delta T}{T}({\bf n'})\rangle_{{\bf n}\cdot{\bf n}'}=\sum_\ell \frac{2\ell+1}{4\pi}C_\ell P_\ell({\bf n}\cdot{\bf n}')
\nonumber
\end{eqnarray}
where $\langle \cdots \rangle$ is the ensemble average and $C_\ell$ is the angular power spectrum of fluctuations.
We can calculate $\frac{\Delta T}{T}$ by evolving the Boltzman equation for the radiation distrubution function,
coupled to the perturbed field equations presented above.

\begin{figure}
\epsfig{file=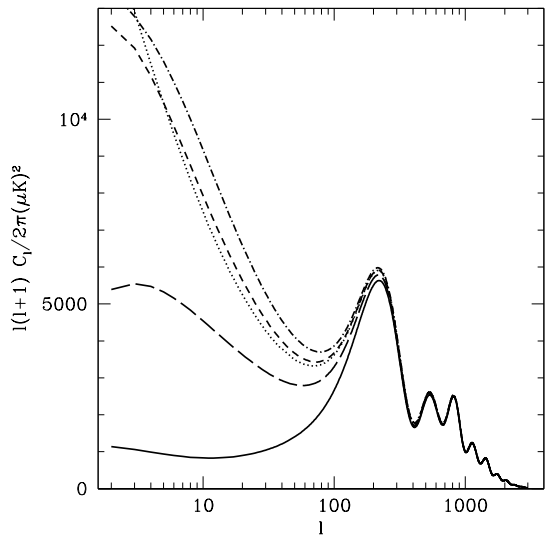,width=3.in}
\caption{The Cosmic Microwave Background angular power spectrum $C_\ell$ for the models described in the realistic background evolution.
The solid curve is model A (the $\Lambda$EBI model) which is indistingushable from the best-fit WMAP-5 $\Lambda$CDM model.
The long-dash curve is model B, the short-dash curve is model C, the dot-dash curve is model D  and the dotted curve is model E.
All models have the same initial amplitude, same tilt ($0.962$)
and non-zero optical depth to reionization ($0.088$) as the best-fit WMAP5 model.
}
\label{fig_cl}
\end{figure}

\begin{figure}
\epsfig{file=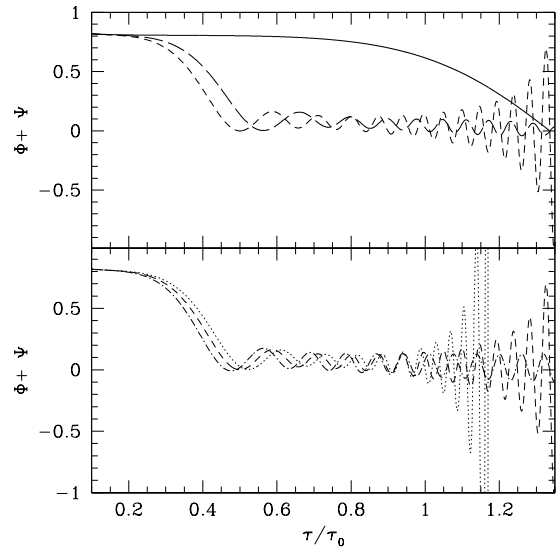,width=3.in}
\caption{The Newtonian potential combination $\Phi + \Psi$  which is relevant to the integrated Sachs-Wolfe effect
for the same set of models A-E, plotted against $\tau/\tau_0$ where $\tau_0$ is the conformal time today.
In the upper panel we display model A (solid curve),
model B(long-dash curve) and model C (short-dash curve).
In the lower panel we show again model C (short-dash curve) to be compared with
model D (dot-dash curve) and model E (dotted curve).
Notice that $\Phi+\Psi$ for models B-E oscillates during the transition to deSitter phase (which is usually a constant-w phase),
while models C-E also start to diverse during the deSitter phase. The presence of a bare cosmological constant in model B
seems to curb the divergence, although the oscillation remains.
}
\label{fig_potentials}
\end{figure}

As suggested above, the evolution of EBI field as dark matter is exactly equivalent to that of ordinary dark matter. Hence its
effect on the CMB will be equivalent and we therefore expect that such observables as the peak positions and heights will
be preserved. This is clearly so if we look at the solid curve in Figure \ref{fig_cl}- it is indistiguishable from $\Lambda$CDM.
Severe differences can arise depending on how the EBI field evolves in the accelerating era. If EBI continues to evolve as
dark matter then the evolution of the gravitational potentials is such that, again, the angular power spectrum is indistinguishable
from that of $\Lambda$CDM. This is clearly not so in the case where the EBI field drives acceleration. The large scale behaviour
of the $C_\ell$s is strongly dependent on the integrated Sachs Wolfe effect which is roughly given by:
\begin{eqnarray}
(\frac{\Delta T}{T})({\bf n})\simeq \int_{\tau_*}^{\tau_0}d\tau (\Phi'+\Psi')[(\tau,{\bf n}(\tau_0-\tau)]
\end{eqnarray}
where primes, $'$, are derivatives with respect to conformal time, $\tau_0$ is conformal time today and $\tau_*$ is conformal time at recombination.
Note that the combination of potentials is the same as presented in equation (\ref{isw_pot}) and its evolutions is clearly
different from the one experienced in the $\Lambda$EBI case. Indeed in Figure \ref{fig_potentials} we plot the evolution of
$\Phi+\Psi$ for a few cases labeled in Figure \ref{equal_dist}. Quite clearly the unstable, oscillatory behaviour is triggered early on
and hence we expect it affect relatively small scales. This is clear from looking at Figure \ref{fig_cl} where the modifications
to the $C_\ell$s, through the integrated Sachs-Wolfe is present all the way to $\ell\sim 150$, well into the first peak. Interestingly enough, for the phantom case, the accelerating phase kicks in later
 and hence there is a smaller integrated Sachs-Wolfe effect for
$\ell>10$; once acceleration kicks in, however, it is much stronger than in the other cases and has a dramatic effect on the largest
scales of the $C_\ell$s.

\begin{figure}
\epsfig{file=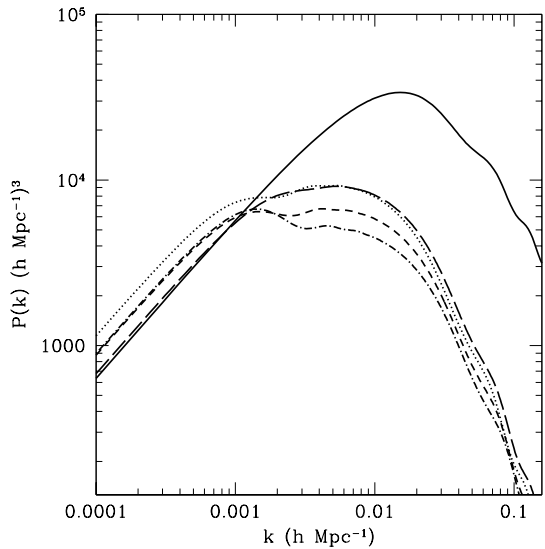,width=3.in}
\caption{The baryon power spectrum $P(k)$ for the same set of models A-E.
Once again we show model A (solid curve) which is indistingushable from the best-fit WMAP-5 $\Lambda$CDM model.
The long-dash curve is model B, the short-dash curve is model C, the dot-dash curve is model D  and the dotted curve is model E.
All models have the same  tilt ($0.962$) as the best-fit WMAP5 model.
}
\label{fig_pk}
\end{figure}

\begin{figure}
\epsfig{file=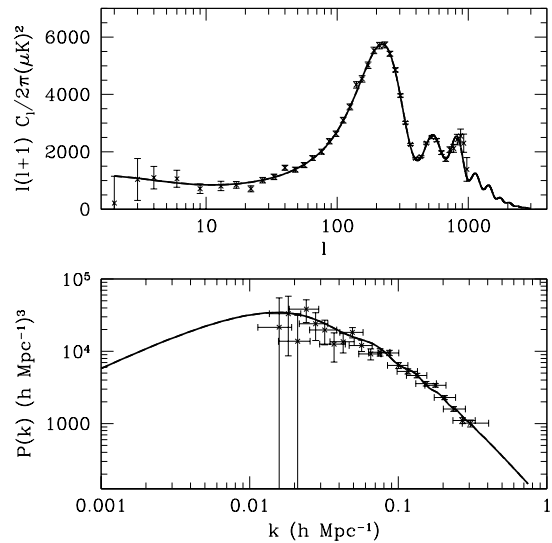,width=3.in}
\caption{The  Cosmic Microwave Background angular power spectrum $C_\ell$ (upper panel) with WMAP-5 data and
baryon power spectrum $P(k)$ (lower panel) for the $\Lambda$EBI model with SDSS data (model A).
Both spectra are indistingushable from the best-fit WMAP-5 $\Lambda$CDM model.
}
\label{fig_cl_pk}
\end{figure}

In the same way, we can directly relate the fluctuations in the galaxy distribution directly to the density contrast. It is convenient to look at the power spectrum of the density fluctuations by taking the Fourier transform of $\delta({\bf k})$, where
${\bf k}$ is the wave number and
constructing the variance:
\begin{eqnarray}
P(k)=\langle |\delta({\bf k})|^2\rangle \nonumber
\end{eqnarray}
Once again, we find that that the $\Lambda$ EBI model looks identical to a $\Lambda$CDM model. In Figure \ref{fig_pk} we plot
such a model with a choice of parameters that render it indistinguishable from the best-fit WMAP-5 $\Lambda$CDM model. It is
clear that this is not true of the EBI model where we find that there is a strong shortage of power on small scales as well as
a much broader turnover associated to the radiation matter transition. The effect is sufficiently dramatic that we don't even attempt to
compare the EBI  model to the angular power spectrum as measured by WMAP-5 \cite{WMAP} or the power spectrum of galaxy fluctuations as
measured by the Sloan Digital Sky Survey \cite{SDSS} in Figure \ref{fig_cl_pk}- the EBI model is not a viable candidate for a theory of
structure formation- while the $\Lambda$EBI model is quite clearly a good candidate.

\begin{figure}
\epsfig{file=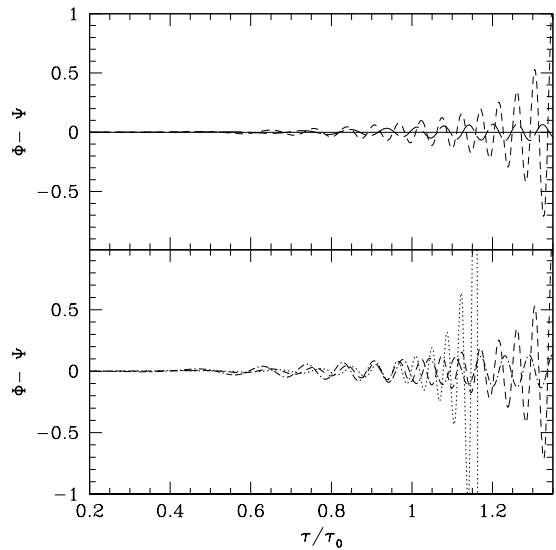,width=3.in}
\caption{The Newtonian potential combination $\Phi - \Psi$
for the same set of models A-E, plotted against $\tau/\tau_0$ where $\tau_0$ is the conformal time today.
In the upper panel we display model A (solid curve),
model B(long-dash curve) and model C (short-dash curve).
In the lower panel we show again model C (short-dash curve) to be compared with
model D (dot-dash curve) and model E (dotted curve).
Notice that $\Phi-\Psi$ for models B-E oscillates during the transition to deSitter phase (which is usually a constant-w phase),
while models C-E also start to diverse during the deSitter phase. The presence of a bare cosmological constant in model B
seems to curb the divergence, although the oscillation remains.
}
\label{fig_potentials_minus}
\end{figure}

The evolution of perturbations in the EBI model do have an interesting feature that is worth noting. In theories of pressureless dark matter, such
as WIMPs, the evolution of perturbations is such that the two gravitational potentials are effectively identical, i.e. $\Phi=\Psi$. It has
been pointed out that in many, if not all, theories of modified gravity, these potentials will differ from each other \cite{bertschinger}
and that this may be a smoking gun for modified theories of gravity. A plethora of observational techniques have been proposed,
cross correlating galaxy surveys with weak lensing surveys and with measurements of the CMB \cite{bean} and is one of the main
science targets of up and coming experiments such as the Euclid project. As pointed out in \cite{kunz} such a signature
is not exclusive to modified gravity and it suffices that the dark sector have a component that takes the form of anisotropic
stress. This is indeed what we find in this theory and specifically in the case of EBI, where $S$ and $\Pi$ can have a substantial
effect on the evolution of perturbations. In Figure \ref{fig_potentials_minus} we illustrate this fact by plotting the evolution of $\Phi-\Psi$
for a selection of models. Granted that we have been unable to find an EBI model that fits the data and the question still remains whether
it is still possible to constrain a fundamental theory of dark matter with $\Phi-\Psi$.

\section{Discussion}
In this paper we have explored the cosmology of Universe permeated by a field that obeys the Eddington-Born-Infeld
equation. As shown in \cite{banados2}, such a field can play the dual role of dark matter and dark energy and
therefore supplies us with a counterpart to the Chaplygin gas as a possible unification of the dark sector. In
our analysis we have shown that there are other regimes in which the EBI field can play a different
role, either as an alternative to simply dark matter or as a source of energy that can renormalize the
cosmological constant.

We have then looked at the effect the EBI field has on the growth of structure. We show that it can
be described in terms of a set of fluid variables, akin to the construction of \cite{Hu} and then
identify the different key regimes. During the dark matter dominated regime, i.e. the regime in
which $w_\E\simeq0$ and the EBI field dominates, the evolution of perturbation is exactly as in
the standard scenario in which the dark matter field is described by massive, non-relativistic particles. The gravitational potentials are constant and indistinguishable during this era, under the assumption that the initial shear and entropy is negligible.
Distinctive signatures emerge in a period of accelerated expansion. If the EBI field dominates and
is responsible for cosmic acceleration, there is a clear instability in the gravitational potentials; they not only
grow but $\Phi+\Psi$ diverges leading very rapidly to an overwhelming integrated Sachs-Wolfe effect on
large scales. It is difficult to reconcile the angular power spectrum of fluctuations and the power spectrum of
the galaxy distribution predicted by an EBI theory which unifies the dark sector, with current data. If we restrict
ourselves to a regime in which the EBI field simply behaves as dark matter, then, as expected, we find the
our best fit model to be entirely indistinguishable from the standard, $\Lambda$CDM scenario.

The EBI field can clearly play an important role in cosmology and, in particular, as a non-particulate
form of dark matter. It's interpretation becomes interesting if we view the theory as bi-metric and
$q_{\mu\nu}$ as the true, geometric, metric of space-time; it is then this metric which is interpreted as dark matter.
This is the point of view implicit in \cite{banados1}.
What we mean by the "true" metric of space-time is of course open to debate. Clocks and rulers will
feel $g_{\mu\nu}$ and hence real geometry will be built out of it and in this case $q_{\mu\nu}$ plays a purely
auxiliary role as an extra field.

We would like to re-emphasize that the generalized EBI model, in which the EBI field drives cosmic acceleration, gives us an interesting example
of a theory with an exotic signature: the mismatch between $\Phi$ and $\Psi$ . A number of methods have been proposed to do
ease out this mismatch from current and future data sets \cite{bean}. We have
found that the EBI field can source this mismatch without modifying gravity. This is not surprising: the EBI
field is a two tensor with four scalar degrees of freedom. One linear combination of these
degrees of freedom can be seen as anisotropic stress which can freely source the $\Phi-\Psi$. It turns out
that its effect is severe enough that the integrated Sachs-Wolfe effects generated is too extreme to be
reconcilable with current observations. But it does suggest that it may be possible to build models which
don't modify gravity, generate accelerated expansion and could be confused with bona-fide modified theories
of gravity \cite{kunz}.  Consistent parameterized frameworks such as~\cite{ppf} may be able to provide alternative ways
to distinguish such theories and it would be interesting to find the EBI's predictions for these frameworks.

Finally, we would like to point out that the EBI model is a viable alternative to the $\Lambda$CDM
but which may have particular features which make it stand out. As we saw in section \ref{ebicdm}, even though
the evolution of perturbations may be equivalent to that of that CDM if one assume no pressure perturbations
and shear in the initial conditions, the non-linear evolution will be different. Pressure perturbations and shear
will be generated at the non-linear level and may play a significant role in the small scale structure of galaxies
and clusters. Indeed, one of the major problems that $\Lambda$CDM has had to face is the excess of small scale power
compared to observations \cite{small}. $\Lambda$EBI may have a natural dynamical solution to this problem. This is one
of the many aspects of this theory we wish to explore further.

\label{discussion}

{\it Acknowledgments}:
We thank Andy Gomberoff,  Nemanja Kaloper, Davi Rodrigues, and Tom Zlosnik for discissions. Part of this work was undertaken under the auspices of STFC and the Beecroft Institute for Particle Astrophysics and Cosmology. MB was partially supported by Fondecyt (Chile) Grants \# 7080116 and \# 1060648.
Research at the Perimeter Institute is supported in part by NSERC and by the Province of Ontario through MEDT.


\end{document}